\documentclass[12pt,preprint]{aastex}
\usepackage{epsfig}  

\shortauthors{Cui}
\shorttitle{X-ray Flaring Activity of Mrk~421}

\begin{document}

\title{X-ray Flaring Activity of Mrk~421} 

\author{Wei Cui} 
\affil{Department of Physics, Purdue University, West Lafayette, IN 47907; 
cui@physics.purdue.edu}

\begin{abstract}

We report results from a systematic search for X-ray flares from Mrk~421, 
using archival data from the {\em Rossi X-ray Timing Explorer} (RXTE) 
satellite. The flares are clearly seen over a wide range of timescales. 
The quasi-continuous coverage of the source with the All-Sky Monitor (ASM) 
reveals frequent occurrence of major flares that last for months. On a 
few occasions, the source was intensively monitored with the more sensitive 
pointing instruments aboard RXTE. The data from these observations shows 
the presence of X-ray flares of much shorter durations, ranging from weeks 
down to less than an hour. For the first time, we clearly resolved the
sub-hour flares from Mrk 421. Moreover, Fourier analyses reveal 
variability on even shorter timescales, up to about 0.01 Hz. The source 
appears to behave differently 
in its spectral properties during different flares, large or small, which 
is intriguing. While significant hysteresis is observed to be associated 
with spectral evolution in some cases, little is seen in other cases. 
Sometimes, 
the shape of the X-ray spectrum hardly varies across a flare. Therefore, 
the phenomenology is complex. The observed hierarchical structure of the 
X-ray flares seems to imply the scale-invariant nature of the phenomenon, 
perhaps similar to solar flares or rapid X-ray flares observed of 
stellar-mass black holes in this regard. Combined with other results, the 
observed 
flaring timescales seriously constrain the physical properties of X-ray 
emitting regions in the jets of Mrk~421. 
 
\end{abstract}


\section{Introduction}

Mrk~421 is one of the brightest extragalactic X-ray sources in the sky. It
belongs to the group of BL Lac objects, a sub-class of radio-loud active 
galactic nuclei (AGN). The BL Lac objects are known to be highly variable 
at nearly all wavelengths and their spectral energy distributions (SEDs) 
are thought to be dominated by radiation from relativistic particles in the 
jets that propagate roughly along the line of sight (Urry \& Padovani 1995). 
Mrk~421 is also one of the few BL Lac objects that were discovered to emit 
strongly at TeV energies in the past decade (Punch et al. 1992). In fact,
the total energy output of Mrk~421 in gamma-rays seems comparable to that 
in X-rays (Buckley et al. 1996; Maraschi et al. 1999; Krawczynski et al. 
2001). The SED of the source is dominated by the X-ray and gamma-ray 
emission. While the origin of gamma-ray photons is still being debated, it 
is generally agreed that X-ray photons originate mostly in the synchrotron 
radiation from highly relativistic electrons in the jets.

BL Lac objects are known for their flaring activities. X-ray flares with 
duration of longer than a day have been seen from Mrk~421 (Tanihata et al. 
2001; Fossati et al. 2000; Brinkmann et al. 2003). Remarkably, a strong 
flare was detected from
the source at TeV energies that lasted only for about an hour (Gaidos et 
al. 1996). The TeV flare showed significant sub-structures that that were
of even shorter durations. The origin of X-ray or TeV flares is hardly
understood. The 
flares are often thought to be related to internal shocks in the jets of a 
blazar (Rees 1978; Spada et al. 2001), or to major ejection events of new 
components of relativistic plasma into the jet (e.g., B\"ottcher et al. 
1997; Mastichiadis \& Kirk 1997). Despite such uncertainty, the measured 
short flaring timescale alone has already severely constrained the size 
of the TeV emission region and the Doppler factor\footnote{Defined as
$\delta \equiv \Gamma^{-1} (1-\beta cos \theta)^{-1}$, where $\Gamma$ and 
$\beta$ are the Lorentz factor and speed (in units of the speed of 
light $c$) of the bulk motion, and $\theta$ is the angle to the 
line-of-sight} of the jets (Gaidos et 
al. 1996), as well as the accretion process in the system (Celotti et al. 
1998). If X-ray and TeV photons 
are associated with the same population of emitting electrons, as often 
postulated based on the observed correlated behaviors of the source in the 
two energy bands (e.g., Maraschi et al. 1999), X-ray flares of similarly 
short duration should also occur. In this context, we conducted a 
systematic search for such flares, 
making use of the rich database on Mrk~421 provided by the {\em Rossi 
X-ray Timing Explorer} (RXTE).

\section{Data}

Mrk~421 is one of the X-ray sources that have been frequently observed with
RXTE over the past 8+ years. It is certainly the most observed blazar. The 
RXTE observations often represent X-ray coverage of Mrk~421 in a 
multi-wavelength campaign. Archival data is now available from campaigns 
conducted in 1996 (under the Guest-Observing programs 10341 and 10345), 
1997 (20341), 1998 (30261, 30262, and 30269), 2000 (40182), and 2001 (50190 
and 60145). The great majority of the observations were made in ``snap-shot'' 
modes, with effective exposure times ranging from half a kilosecond to several 
kiloseconds. Deeper observations (10-20 ks) were occasionally carried out.
In the following, we briefly describe the RXTE instruments and the data 
reduction procedures adopted for this work.

\subsection{Instruments}

RXTE carries on board two co-aligned, large-area detectors (Bradt, Rothschild, 
\& Swank 1993): the Proportional Counter Array (PCA) and the High-Energy X-ray 
Timing Experiment (HEXTE), in addition to the All-Sky Monitor (ASM). Both of
the pointing instruments have a 1\arcdeg\ field-of-view. The PCA consists 
of five nearly identical proportional counter units (PCUs). It has a total 
collecting area of about 6500 $cm^2$ and covers a nominal energy range of 
2--60 keV. However, the operational constraints often require that some of the 
PCUs be turned off. Exactly which PCUs are off varies from observation to 
observation, but PCU 0 and PCU 2 are nearly always in operation. Further 
complication (to data analysis) developed following the loss of the front veto 
layer in PCU 0. The result is that the data from PCU 0 is more prone to 
contamination by events caused by low-energy electrons entering the detector. 
This is particularly relevant to the study of variability of weak sources like 
Mrk~421.  

Although the HEXTE covers a very interesting energy range (nominally 15--250
keV) for AGN studies, it is not very sensitive. We did analyze all of the 
HEXTE data for this work, but found it not very useful for this work, since 
we are mostly interested in detecting rapid X-ray flares from Mrk~421. The 
HEXTE results will be presented in a future paper, along with other spectral
results, on broad-band SED variability. We will not discuss them any further 
here.

The ASM consists of three proportional counters, each of which has a 
6\arcdeg $\times$ 90\arcdeg\ field-of-view, and covers about 80\% of
the sky upon the completion of one full rotation, which takes about 1.5 
hours. However, the orientation of the rotation axis of the ASM is fixed 
by the pointing configuration of the PCA and HEXTE, so there is generally
no flexibility in controlling which 80\% of the  sky is covered at a given 
time. This is a major source of data gaps in the 
quasi-continuous light curve of a particular source, in addition to  
Earth occultation and proximity of the source to the sun. The ASM scans the 
sky in 6\arcdeg\ steps, which are often referred to as ASM dwells. The 
exposure time of each dwell is 90 seconds. 

\subsection{Data Reduction}

Multiple data modes are usually employed in a PCA observation. Two of the
available data modes, {\em Standard1} and {\em Standard2}, are always run
and are thus referred to as the standard modes. We rely almost exclusively
on the {\em Standard2} data for this work. The data has a time resolution
of 16 seconds, which is sufficient for our purposes here, and covers the 
entire PCA passing band with 128 energy channels.

The PCA data was reduced with the latest version of {\em FTOOLS} (v5.2) that 
is distributed as part of the software suite {\em HEASOFT} 
(v5.2).\footnote{see http://heasarc.gsfc.nasa.gov/docs/software/lheasoft} 
For an observation, we first filtered data by following the standard procedure 
for faint sources (see the online RXTE Cook Book\footnote{http://heasarc.gsfc.nasa.gov/docs/xte/recipes/cook\_book.html}), which resulted in a list of Good 
Time Intervals (GTIs). We then simulated background events for the observation 
using the latest background model that is appropriated for faint sources 
(pca\_bkgd\_cmfaintl7\_eMv20020201.mdl). Using the GTIs, we proceeded to 
extract a 
light curve from the data (combining all active PCUs) in each of the following 
energy bands: 2.0--5.7 keV, 5.7--11 keV, 11--60 keV, and 2.0--60 keV. Note that
the boundaries of each band are matched up as closely as possible across 
different PCA epochs but are only approximate (up to $\pm$0.2 keV). We 
repeated the steps to construct the corresponding background light curves from 
the simulated events. Finally, we subtracted off the background to obtain the 
light curves of the source. In an attempt to avoid possible artifacts caused 
by background variations (see previous section), we excluded data from PCU 0 
when we made light curves from observations conducted in Epoch 5. 

The ASM data is publicly available. The light curves come in three energy 
bands: 1.5--3 keV, 3--5 keV, and 5--12 keV, providing crude spectral 
information about a source. We obtained the light curves of Mrk~421 from the 
MIT archive.\footnote{http://xte.mit.edu/asmlc/srcs/mkn421.html\#data} 
We chose to filter out data points with error bars (on the count rates in the 
summed band) greater than 2.0 c/s, which constitutes about 10\% of the data. 
We then weighted the raw data by $1/\sigma^2$ and rebinned it to produce 
weekly-averaged light curves. 

\section{Results}

To provide a global view of flaring activity of Mrk~421, Fig.~1 shows the ASM
light curve of Mrk~421 for the 1.5--12 keV band over roughly an eight-year 
period. The light curve shows two things very clearly. First of all, the 
source 
is highly variable in X-rays. It goes from below the detection threshold to 
a flux level of 50--60 mCrab in the ASM passing band. Secondly, frequent 
flares are easily identifiable from the 
light curve, with durations ranging from less than a month up to almost a year. 
There appears to have been 4 major flares over 8 years, or one every two years
on average, although the source seems to be more active in recent years. There 
is hardly any time period that resembles a ``quiescent state''. Outside the 
major flares, which we refer to as the ``low state'' for convenience, 
the average flux of Mrk~421 is about 5 mCrab (see Fig.~1). We stress, however, 
that smaller (both in amplitude and duration) flares are nearly always 
present.

To examine possible spectral evolution of Mrk~421 during major flares, we 
computed the ratio between the ASM rates in the 3--12 keV band to those in
the 1.5--3 keV band and used it as a measure of the spectral shape. Fig.~2 
shows the time series of the hardness ratio that covers the periods of the 
two largest flares. The data is quite noisy, even with weekly binning, but
there does not seem to be any strong spectral evolution across either 
flare. 

\subsection{Low State}

Mrk~421 was in an extended low state throughout 1996 and 1997 (see Fig.~1). It 
was monitored regularly by the PCA, with particularly good coverage in 1996, 
and the data allows us to examine, in a more detailed manner, the properties 
of the source in this state. The source is by no means ``quiet'' during this
period, as shown in Fig.~3. With much improved sensitivity of the PCA (as
compared to the ASM), we now see the presence of flares of shorter durations. 
Two larger flares are recognizable from the figure (despite the presence of
data gaps) that last for weeks. Superimposed on them are many ``spikes'' 
whose duration can be as short as about a day. The source reaches the peak
flux of roughly $10^{-9}\mbox { }ergs\mbox{ }cm^{-2}\mbox{ }s^{-1}$ (in the 
2--60 keV band) in this state but can drop below even the PCA detection
threshold ($\lesssim 10^{-11}\mbox { }ergs\mbox{ }cm^{-2}\mbox{ }s^{-1}$). 

As with the ASM data, we defined a hardness ratio (5.7--60 keV/2--5.7 keV) to
examine the spectral behavior of Mrk~421 during these short flares. Though
crude, the hardness ratio does provide a model-independent way of presenting 
the data. In Fig.~4, we plotted the hardness ratio against the overall count 
rate of the source for the two flares separately. While there seems to be a 
common trend of spectral 
hardening as the source brightens, there is also significant difference 
between the two cases. Significant hysteresis appears to be associated with
the spectral evolution during the first flare but little during the second 
one. Similar hysteresis patterns have been seen before (e.g., Takahashi et 
al. 1996; Kataoka et al. 2000; Zhang 2000; Giebels et al. 2002; 
Falcone et al. 2004), and seems
to be be quite common among blazars. The patterns that we have observed here 
are complex in shape, e.g., with embedded sub-loops, perhaps reflecting the 
co-existence of flares on different timescales. Also, both clockwise and
counter-clockwise running sub-loops seem to be present, so the phenomenology
is not simple.

We searched for flares of even shorter duration (less than a day). The effort
was complicated by the presence of data gaps (mostly due to earch occultation 
or South-Atlantic Anomaly crossing) for relatively 
long observations. We failed to detect any 
significant flares on timescales of tens of minutes to hours, although the 
light curve seems to show variation on these timescales in several cases. At 
the level of the systematic variations seen, however, one must be worried 
about imperfect background modeling. To illustrate the point, we show a 
possible flare in Fig.~5, along with the corresponding light curve of 
the background. The variation patterns seem to be quite similar in the two 
cases, which is of concern. In this example, however, the observed amplitude 
of the variation is perhaps too large to be
attributed entirely to the inaccuracy of the background model.

\subsection{Flaring State}

Mrk~421 has been quite active since 1998 (see Fig.~1). The source was 
regularly monitored by the RXTE during each of the major flares, with
comprehensive coverages in 1998 and 2001. In general, the flaring 
activity also becomes more intense on short timescales (less than a day) 
during these periods and, quite remarkably, rapid flares (of duration 
less than about an hour) are often seen. Fig.~6 displays a 
montage of light curves made from a stretch of consecutive observations 
during the 2001 episode. The results are quite typical of the source 
in the flaring state. The figure shows an impressive array of 
variation patterns on a wide range of timescales. Of particular interest
are prominent flares that last for less than a day, which are not as
obvious in the low state. 

To investigate any spectral evolution during the short flares, we show
in Fig.~7 the corresponding hardness--intensity diagrams. Many different 
patterns are observed here. In some cases, the hardness ratio 
increases nearly monotonically with the count rate, while, in some other
cases, the behavior is much more complicated. For example, the panels b 
and c show the instances of strong hysteresis in the spectral evolution 
of the source. To present a more direct view, we expanded Panel c (of 
Fig.~7) and indicated the time progression of the evolution in Fig.~8. 
Overall, the source appears to evolve along clockwise loops in this case. 
Once again, multiple branches (or sub-loops) reflect the co-existence of 
flares on a wide range of scales. Interestingly, in some cases, there is 
hardly any measurable spectral change across a flare.

The light curves in Fig.~6 clearly reveal the presence of rapid flares 
that last for less than an hour. Panel f shows a beautiful example of 
such a flare. In fact, there are
significant sub-structures associated with the event, suggesting
the presence of two overlapping flares of even shorter durations. 
Interestingly, the X-ray spectrum of the source varies little across
this flare (see Panel f in Fig.~7). The effort to precisely determine 
the duration of each flare was, in general, complicated by the presence 
of data gaps, as well as the co-existence of flares on a wide range of 
timescales. Sometimes, only a portion of a flare is seen. The shortest 
rise or decay time seen is about 1000 s, although it can be argued that 
it might be even shorter in some cases (see, e.g., Panel e of Fig.~6).

\subsection{Power Density Spectra}

The observed variability may extend to shorter timescales, on which 
individual X-ray flares become unresolvable. The collective effects of 
such variability can be investigated by adopting a more sophisticated 
time-domain or Fourier-domain based technique. We chose to follow the 
latter approach to obtain a representative power-density spectrum (PDS) 
of Mrk~421 in the low or flaring state. We selected a subset of the 1997 
observations that are relatively long for the low state and, similarly, 
a subset of the 2001 observations for the flaring state. The total
exposure time is comparable for the two data sets.
For each observation, we made a light curve from the {\em Standard1} 
data that has a time resolution of 1/8 s (but no energy resolution). 
We then broke the light curve into segments, each of which is 4096 s 
long (which requires the padding of data gaps or shorter segments with 
the average count rate). We performed Fast-Fourier transformation on 
each segment to obtain a PDS. The PDS was normalized according 
to a scheme proposed by Leahy et al. (1983). The individual PDSs of the
segments were then weighted (by the total number of photons) and 
averaged to obtain the PDS for the observation. To further improve
statistics, we weighted and averaged the PDSs of the selected 
observations in a similar manner. From the resulted PDS, we subtracted 
off noise power due to Poisson counting statistics to obtain the PDS 
of the source. Fig.~9 shows the final PDS for each state. 

The observed PDS can be fitted by a simple power law, $1/f^{\alpha}$, 
where $\alpha=1.9\pm 0.2$ for the low state and $2.26\pm 0.06$ for the 
flaring state, although statistics is quite limited for the low state. 
The latter value is in general 
agreement with the published results for the flaring state (Kataoka et 
al. 2001; Brinkmann et al. 2003). The PDS appears to fall more steeply 
in the flaring state, although the difference is only of marginal
statistical significance. The power-law type of PDS is typical of AGN. 
What is remarkable here is that the variability of Mrk~421 is positively 
detected up to 0.01 Hz (or 100 seconds) in the flaring state! Due to the 
faintness of the source, the low-state data is much noisier. Nevertheless, 
the source variability is clearly seen beyond $10^{-3}$ Hz.

\section{Discussion}

We have observed flaring activities of Mrk~421 on a vast range of timescales
(spanning roughly 4 orders of magnitude). The flares seem to occur at all
times, in the low state or flaring state, and the X-ray variability seen 
might be entirely due to the superposition of these events on different 
timescales. For the first time, we have clearly resolved individual X-ray 
flares that last less than about an hour. They might be similar in 
physical origin to the TeV flare of comparable duration (Gaidos et al. 1996), 
albeit much smaller in amplitude. The only other blazar that is known to 
produce similarly rapid flares is Mrk~501 (Catanese \& Sambruna 2000). 

\subsection{Multi-Scale X-ray Flares}

The seemingly scale-invariant nature of flaring activities in Mrk~421 is 
perhaps best illustrated in Fig.~10. Note the similarity in the occurrence 
of X-ray flares on different timescales. The detectability of 
individual flares on even short timescales might be limited only by the 
statistics of the data. Certainly, for Mrk~421, X-ray variability is 
still measurable on a timescale of 100 s (see Fig.~9). The PDS is well 
described by a power law (although the frequency range is limited). We
note, however, that there appears to be a genuine lack of very rapid flares 
(with duration $<$ 1 hour) in the low state. Only several candidates 
were identified (see Fig.~5 for an example) but their small amplitudes
raise concern about possible influence by background variations.

Similar scale-invariant flaring phenomena  
have been observed of accreting stellar-mass black holes. For instance, the 
X-ray light curves of Cyg X-1 show the presence of flares on timescales of 
months (e.g., Cui, Feng, \& Ertmer 2002) and ``shot'' on much shorter 
timescales (down to milliseconds; Kato, Fukue, \& Mineshige 1998 and 
references therein). It is often thought that the ``shots'' are associated 
with accretion processes in black hole candidates (BHCs), although the 
roles of relativistic outflows have drawn increasingly intense attention in 
recent years, given the ubiquitous presence of jets in such systems. It is 
interesting to note that a 1-ms flare 
from a $10 M_{\odot}$ black hole would roughly correspond to a 1-ks flare 
from a $10^8 M_{\odot}$ black hole, if one scales the dynamical timescales 
by mass and takes into account the observed differences in the jet 
properties between BHCs (with Doppler factors $\delta \sim 1$) and 
blazars ($\delta \sim 10$). The mass of the black hole in Mrk 421 is
estimated to be about $2\times 10^8M_{\odot}$ (Barth, Ho, \& Sargent 2003). 

Further analogy can perhaps be made between blazar flares and solar 
flares. To account for the power-law distribution of the solar flares, 
Lu \& Hamilton (1991) postulated that the coronal magnetic field in the 
Sun is a self-organized critical (SOC) state and that a solar flare is 
the result 
of an avalanche caused by many small magnetic reconnection events. The 
size of a flare is, therefore, only determined by the number of elementary 
reconnect events. In the SOC state, stable regions of all sizes are 
present in the system, so avalanches can occur on all scales (up to the 
size of the system), leading to the observed scale-invariant distribution 
of the flares. Extending the concept to BHCs, Mineshige et al. (1994) 
suggested that accretion disks in these systems might also be in an SOC 
state and that avalanches in the accretion process, triggered by, e.g., 
magnetic reconnection in the disk, could result in the observed 
distribution of ``shots'', as well as the power-law shape of the PDS. 

Lyutikov (2002) discussed the roles of magnetic 
reconnection in the jets of AGN and argued that magnetically dominated 
jets had many advantages over hydrodynamic jets, e.g., in explaining the 
rapid variabilities observed of blazars. Acceleration of particles may
also occur in the reconnection region, and the accelerated particles 
then emit photons at X-ray or gamma-ray energies. It is, therefore,
conceivable that the flaring activity seen in Mrk~421 might originate in 
the same physical process as the solar flares or flares in BHCs. 
Whether the magnetic field in the jets of black holes could be in an SOC 
state remains to be seen. 
Theoretical efforts to quantitatively model blazar variabilities have so 
far been mostly 
concentrated on scenarios that involve internal shocks in the jets to
accelerate particles (e.g., Salvati, Spada, \& Pacini 1998; Protheroe 2002; 
Tanihata et al. 2003). The results are, in fact, quite encouraging. From
the observational point of view, the differences between reconnection-based 
models and shock-based models are expected to be subtle, as discussed by 
Lyutikov (2002), so they will remain unresolved until better data becomes
available. On the other hand, both types of models may also be applicable 
to other types of systems, including gamma-ray bursts (Lyutikov, Pariev, 
\& Blandford 2003) and perhaps also micro-quasars. 
Insights might be gained by looking at these physically similar systems 
collectively.

\subsection{Physical Constraints}

Despite the uncertainties, much can still be learned, in a relatively
model-independent manner, about the physical properties of X-ray emitting
regions from the observed X-ray variability. First of all, all of the 
resolvable flares seen from Mrk~421 appear to be symmetric in shape 
(but see Fig.~5, if the flare is real). This implies that neither the 
synchrotron cooling time nor the particle acceleration time is probably
a major factor in determining the duration of a flare (Protheroe 2002).
Therefore, the observed distribution of the durations of the flares 
may simply reflect that of the sizes of the X-ray emitting regions, if
the (jet-frame) timescale of intrinsic variability is negligible compared
to the light-crossing time. The characteristic (jet-frame) dimension of 
such a region is then given by
\begin{equation}
 l \approx c t_{flare} \delta/(1+z) \approx 10^{15}\mbox{ }cm\mbox{ }
t_{flare,hr} \delta_1, 
\end{equation}
where $t_{flare,hr}$ is the flare duration in units of hours, 
$\delta=10\delta_1$, and $z$ is the redshift of Mrk 421 ($z = 0.031$). 
Given the fact that Mrk~421 is 
nearly always detectable at TeV energies, a large value of $\delta$ 
seems to be always required to sufficiently reduce the photon-photon 
pair-production opacity in the TeV emitting region (to allow TeV photons 
to escape). For instance, Gaidos et al. (1996) showed that the short 
duration of the observed TeV flare would imply $\delta \gtrsim 10$. 
Also, results from modeling the broadband SED of the source show the need 
for large $\delta$ (e.g., Maraschi et al. 1999; Krawczynski et al. 2001), 
very large in some cases ($>$ 50; e.g., Krawczynski et al. 2001). However, 
Georganopoulos \& Kazanas (2003) argued that such extreme $\delta$ values 
would not be required if the jets actually decelerate on sub-parsec scales.
In any case, Eq. 1 implies that the size of the emission regions probably 
ranges from $10^{15}\mbox{ }cm$ to $10^{18-19}\mbox{ }cm$. However, the 
equation might not hold for short-duration flares, because roughly 
symmetric flare profiles could also arise from situations where 
the synchrotron cooling time dominates {\em and} the intrinsic variability 
timescale is much longer than the acceleration time (Kirk \& Mastichiadis 
1999). In this case, the rise or decay time of a flare would be 
approximately equal to the synchrotron cooling time (in observer's frame).

Secondly, the observed decay time of a flare sets a firm upper limit on the 
synchrotron cooling time of the emitting electrons. The synchrotron cooling 
time is given by $ \tau_{syn} \approx 6 \pi m_e c/\sigma_T \gamma_p B^2$ 
(Rybicki \& Lightman 1979), where $m_e$ is the electron rest mass, $\sigma_T$
is the Thomson cross section, $B$ is the magnetic field in the region, and 
$\gamma_p$ is chosen to be the characteristic Lorentz factor of those 
electrons that contribute to the bulk of the observe X-ray emission (at
$E_{p}\sim 10$ keV, where the synchrotron peak of the SED lies). Transformed 
to the observer's frame, this time needs to be less 
than the observed decay time of an X-ray flare, i.e., 
$\tau_{syn}/\delta < t_d$, 
where $t_d$ is the decay time. This requirement leads to a lower limit on 
the magnetic field,
\begin{equation}
B > 0.88 \mbox{ }G\mbox{ }t_{d,3}^{-1/2} \delta_1^{-1/2} \gamma_{p,5}^{-1/2}, 
\end{equation}
where $t_d=10^3t_{d,3}$ s and $\gamma_p = 10^5\gamma_{p,5}$ . Also, the 
observed energies of synchrotron photons are given by
$E_{p}=\delta h\nu_c \equiv (3 e h/4 \pi m_e c) \delta \gamma_{p}^2 B$ 
(Rybicki \& Lightman 1979),
so we have
$B \approx 5.8 E_{p,1} \delta_1^{-1} \gamma_{p,5}^{-2} \mbox{ }G$, 
where $E_p = 10 E_{p,1}$ keV. 
Substituting $B$ in Eq. 2, we derived an {\em upper} limit on the Doppler 
factor,
\begin{equation}
\delta < 16 t_{d,3} E_{p,1}^2 (\gamma_{p,5}/3)^{-3}.
\end{equation}
It is worth noting that $\gamma_{p}$ derived from the synchrotron self-Compton 
description of the observed broad-band SED is typically a few~$\times 10^5$
(e.g., Maraschi et al. 1999; Krawczynski et al. 2001).

\subsection{Spectral Hysteresis}

We observed hysteresis associated with spectral evolution across a flare
in some cases, but not always. For instance, the spectrum of Mrk~421 
varied little across one of the rapid flares (see Panel f in Fig.~6). 
In comparison, Mrk~501 showed significant spectral variation during
rapid flares (Catanese \& Sambruna 2000). In that case, the 
hardness-intensity diagram shows that the source followed a 
counter-clockwise hysteresis loop (though details are more complicated).

Theoretical interpretation of the observed spectral hysteresis is not 
entirely clear at present. Kirk \& Mastichiadis (1999) showed, using
a simplified internal shock model, that the hysteresis could be caused by 
the interplay of three characteristic times associated with synchrotron 
cooling ($\tau_{syn}$), particle acceleration ($\tau_{acc}$), and 
intrinsic variability ($\tau_{var}$), respectively. Two of the four
cases that they discussed might be most relevant to what we have 
observed in this work: (1) $\tau_{var} \gg \tau_{syn} \gg \tau_{acc}$ and 
(2) $\tau_{var} \approx \tau_{syn} \approx \tau_{acc}$. In the former scenario,
the shape of the spectrum does not change and the intensity varies on
the timescale $\tau_{var}$ (which determines the flare duration $t_{flare}$). 
This would explain the lack of spectral evolution during some of the flares 
seen. In the latter scenario, spectral evolution is expected. Moreover, the
evolution should follow a counter-clockwise hysteresis pattern in the
hardness-intensity representation, since changes propagate from 
low energies to high energies (Kirk \& Mastichiadis 1999). Such patterns
were also observed (see Fig.~4). On the other hand, the pattern shown in 
Fig. 8 is mostly clockwise, which would require 
$\tau_{syn} \gg \tau_{var} \gg 
\tau_{acc}$ or $\tau_{syn} \gg \tau_{acc} \gg \tau_{var}$ (Kirk \& 
Mastichiadis 1999; also see Dermer 1998). The latter seems quite unlikely
given the roughly symmetric profiles observed of the flares. 

However, Li \& Kusunose (2000) pointed out that it would be difficult
to draw any firm conclusions based on the sense of a hysteresis loop 
because it is quite sensitive to various parameters in the model, such 
as the overall injection energy. By treating the Comptonization process 
in a consistent manner, they further showed that a hysteresis loop may 
change its sense going from the synchrotron-dominated regime (at low 
energies) to Compton-dominated regime (at high energies). This was confirmed
by B\"ottcher \& Chiang (2002). This is troublesome, from the observational
point of view, because it implies that the choice of energy bands (which
is usually done without {\em a priori} knowledge about the overall SED) 
may affect what we see observationally. To further complicate the 
situation, it is nearly impossible, observationally, to cleanly isolate 
a flare from the hierarchical structure.

\acknowledgments
We wish to thank Mark Ertmer and Kie Li for assistance in data reduction, 
Feng Yuan and Markus B\"ottcher for useful discussions and comments on 
the manuscript. This 
research has made use of data obtained through the High Energy Astrophysics 
Science 
Archive Research Center Online Service, provided by the NASA/Goddard Space 
Flight Center. This work was supported in part by the NASA grant NAG5-13736.

\clearpage

\begin{figure}
\psfig{figure=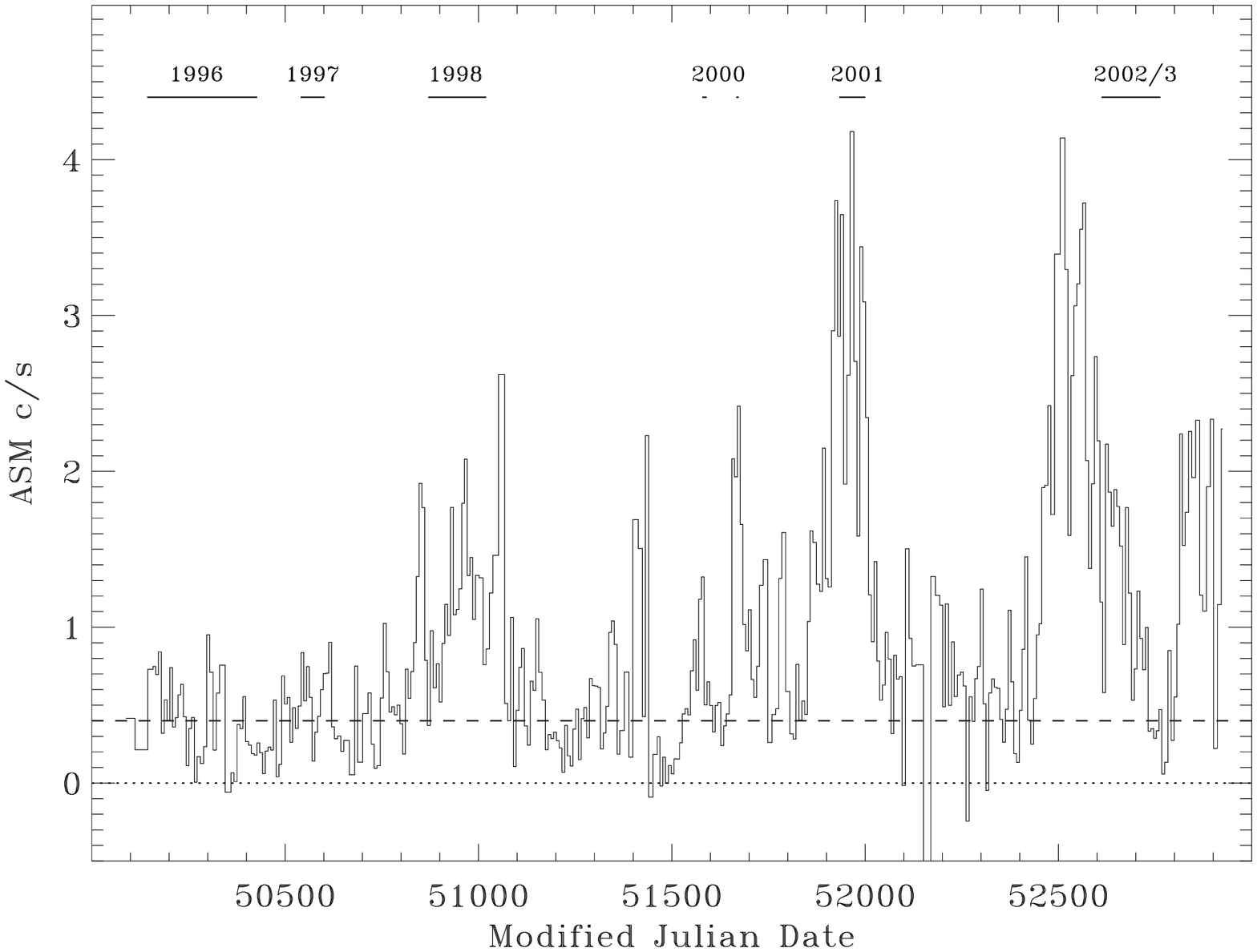,width=5in,angle=90}
\caption{ASM light curve of Mrk~421. Each data point represents a weighted
average of raw count rates over 7 consecutive days (following the exclusion 
of ``bad data''; see text). A few negative data point still remain, which 
are obviously artifacts. For clarity, the error bars are not shown. The 
dashed-line shows roughly the average count rate of the source in the 
low state. The periods of monitoring campaign on Mrk~421 with the 
pointing instruments aboard RXTE are indicated by horizonal lines at the 
top. }
\end{figure}
\begin{figure}
\psfig{figure=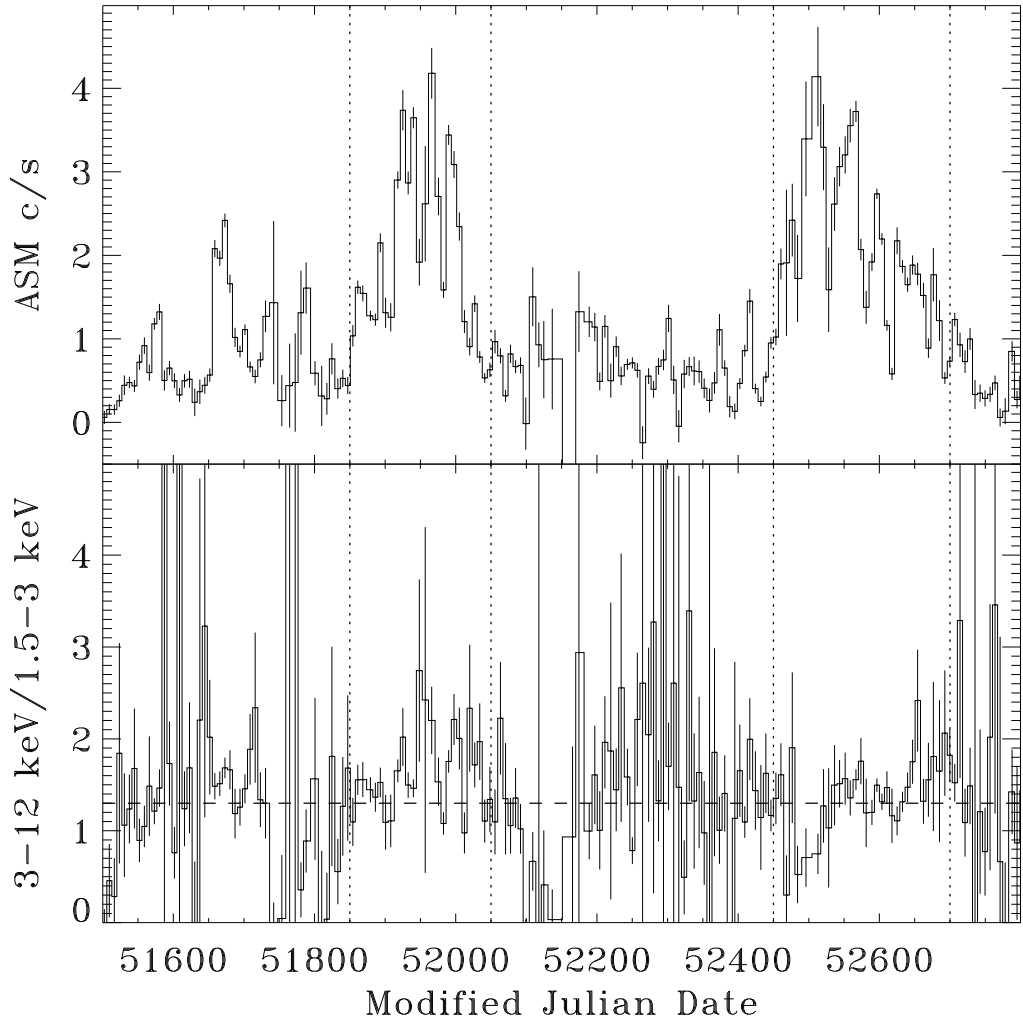,width=6in}
\caption{Spectral variation across major X-ray flares. Two of the flares 
in Fig.~1 are highlighted here. The time series of the hardness ratio in
the bottom panel was derived from the weekly averaged light curves in the
two energy bands. The dashed-line shows roughly the nominal value for the 
low state. }
\end{figure}
\begin{figure}
\psfig{figure=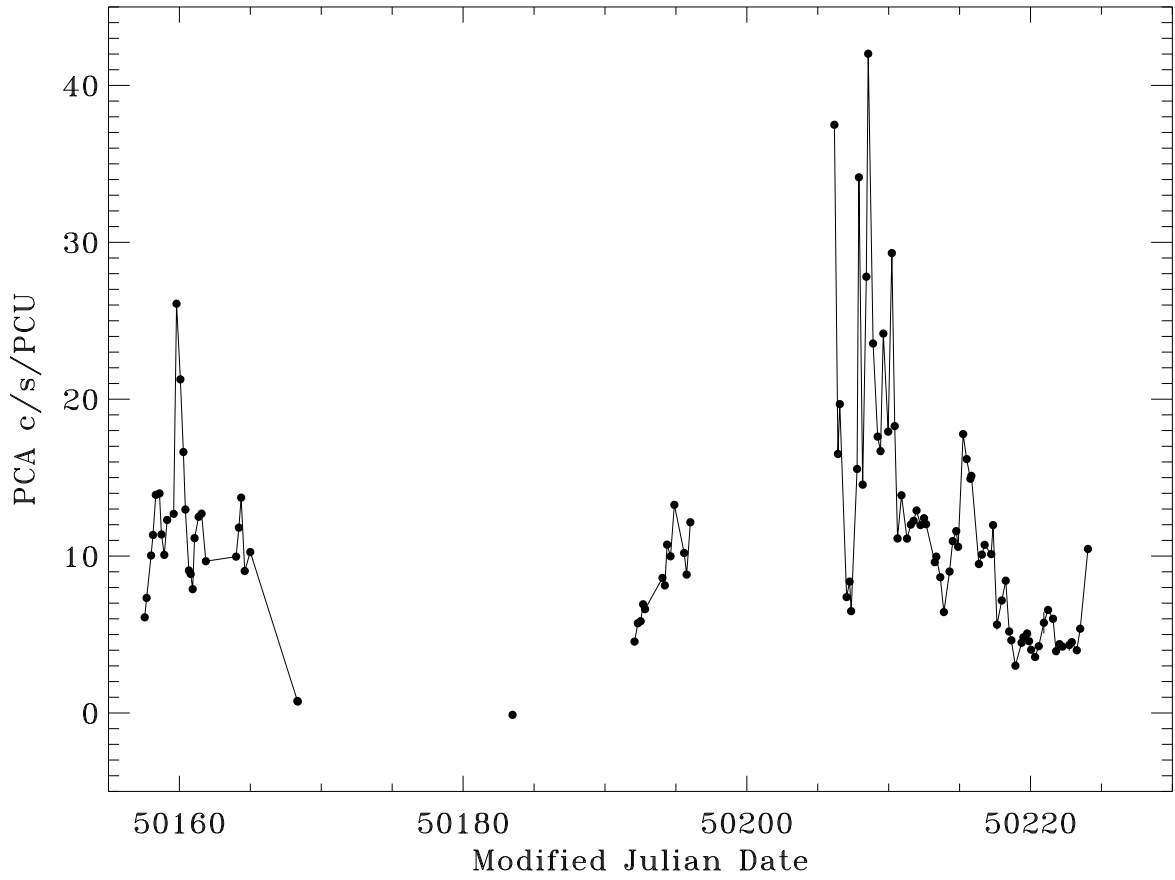,width=5in,angle=90}
\caption{PCA light curve of Mrk~421 in the low state. Only results 
from the 1996 campaign are shown here. The count rates were computed 
for the 2--60 keV band and were averaged with 2048 s time bins. }
\end{figure}
\begin{figure}
\psfig{figure=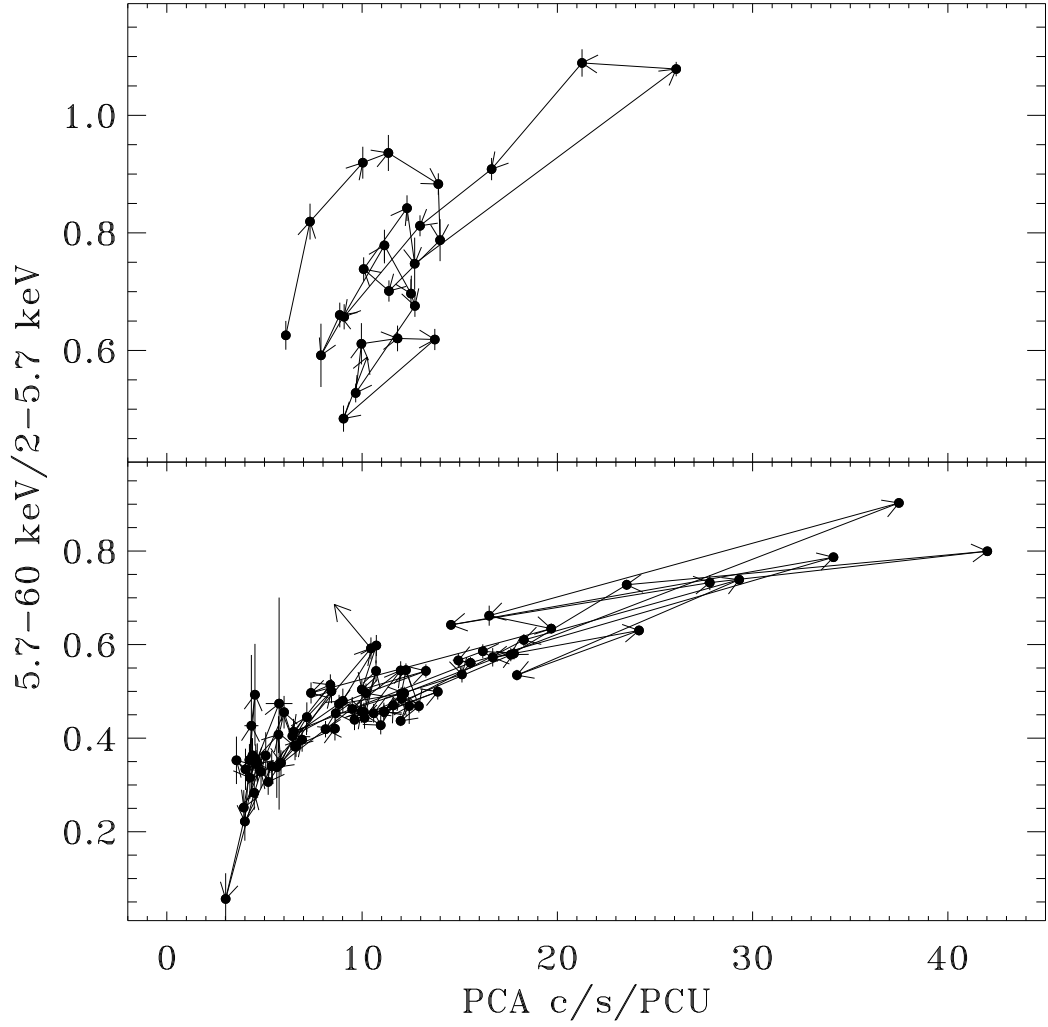,width=6in}
\caption{Spectral variation across short X-ray flares in the low state. 
The results are shown separately for the two flare shown in Fig.~3. The 
arrows indicate the time sequence across a flare. }
\end{figure}
\begin{figure}
\psfig{figure=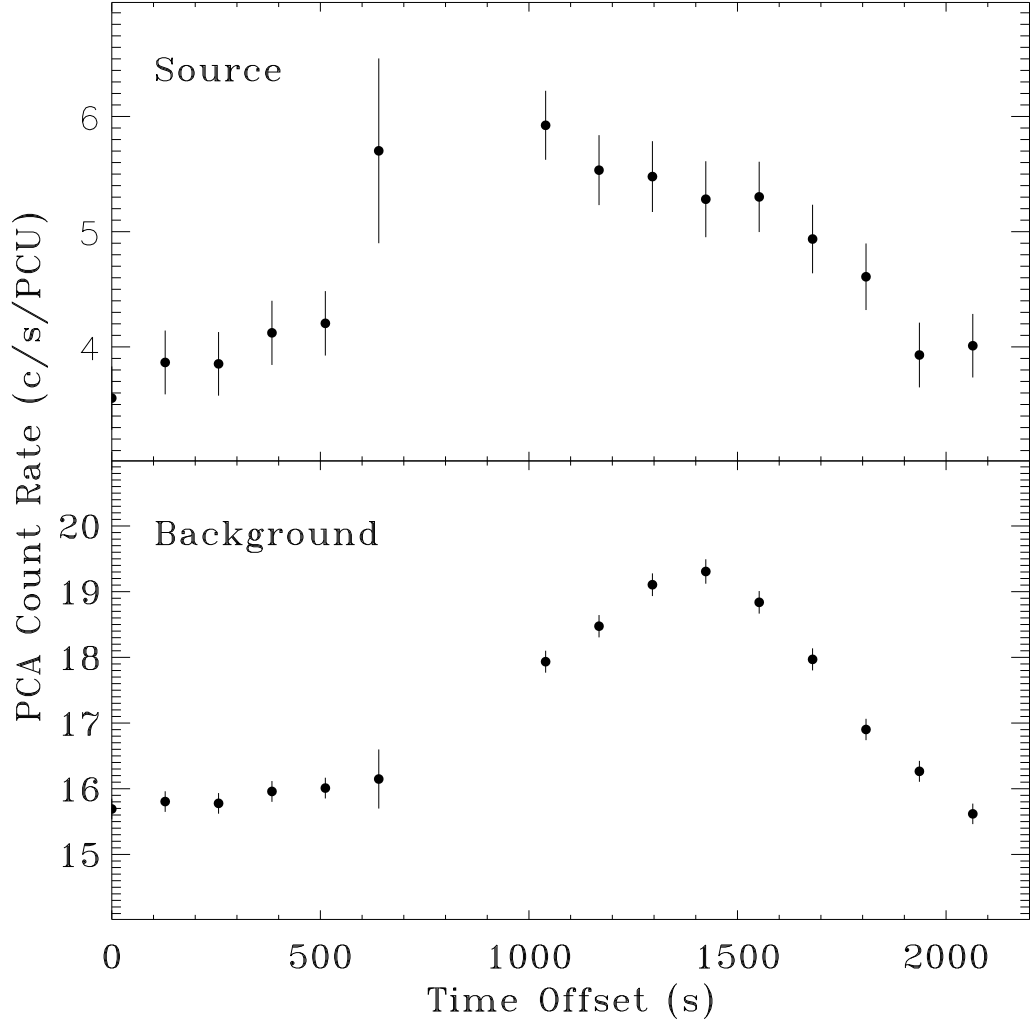,width=6in}
\caption{Possible rapid flare in the low state. The data was taken from one 
of the 1996 observations (10341-02-01-00) and was rebinned to 128 s. The top 
panel shows a {\em background-subtracted} light curve of Mrk~421, while the 
bottom panel shows the background light curve itself for comparison. Note 
the similarity in the variation profiles between the two. }
\end{figure}
\begin{figure}
\psfig{figure=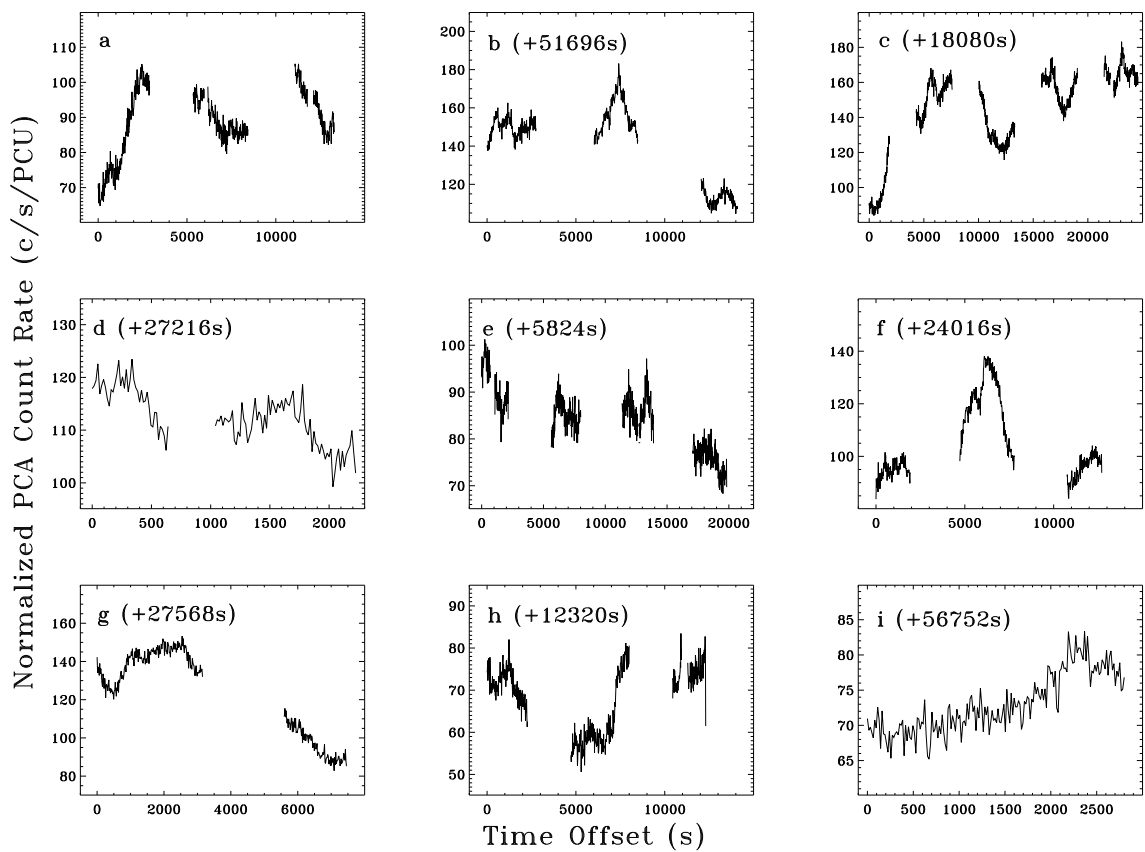,width=4.5in,angle=90}
\caption{Sample light curves of Mrk~421 in the flaring state. They were made 
from a stretch of consecutive observations conducted in 2001. The time is
measured from the beginning of each observation. Also indicated are time 
offsets between the starting times of adjacent observations. }
\end{figure}
\begin{figure}
\psfig{figure=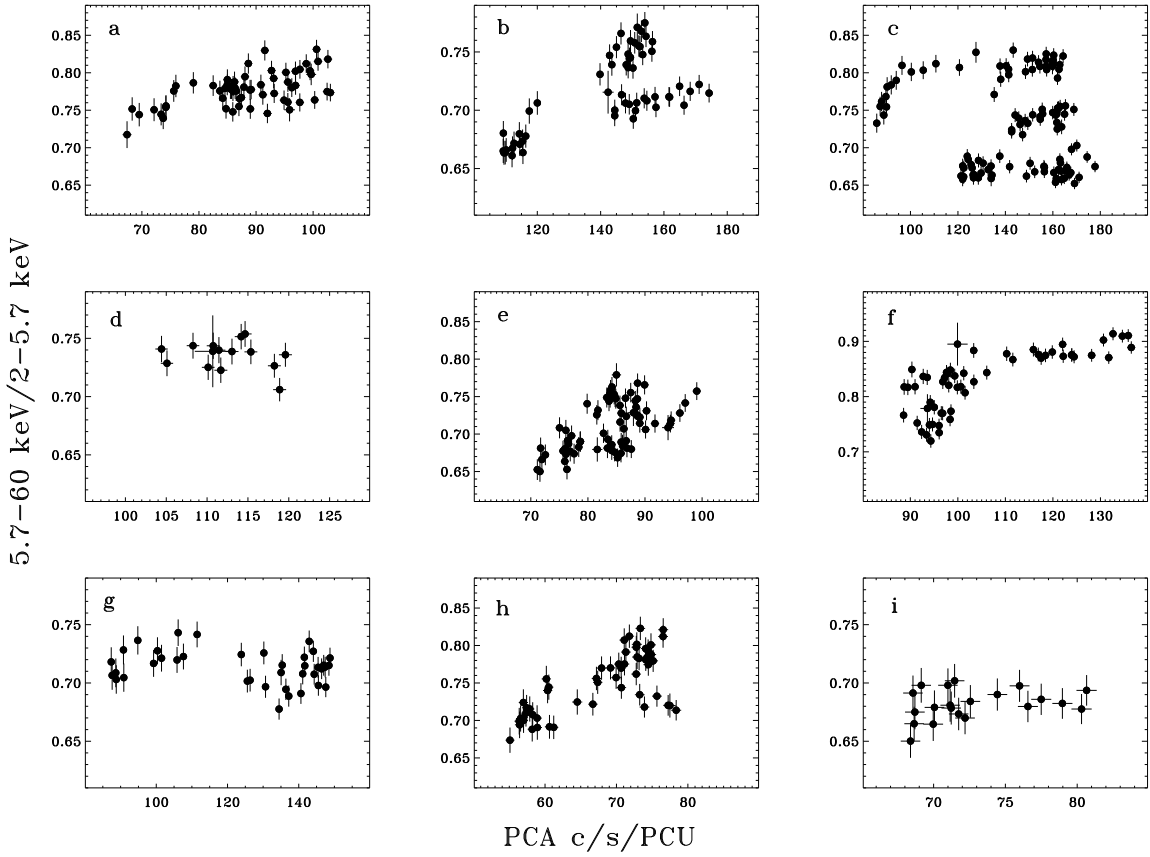,width=4.5in,angle=90}
\caption{Relationship between spectral hardness and count rate. Each panel 
is labeled the same way as in Fig.~6 for cross references. The data was
rebinned to 128 s for clarity. Note the variety of different patterns 
observed. }
\end{figure}
\begin{figure}
\psfig{figure=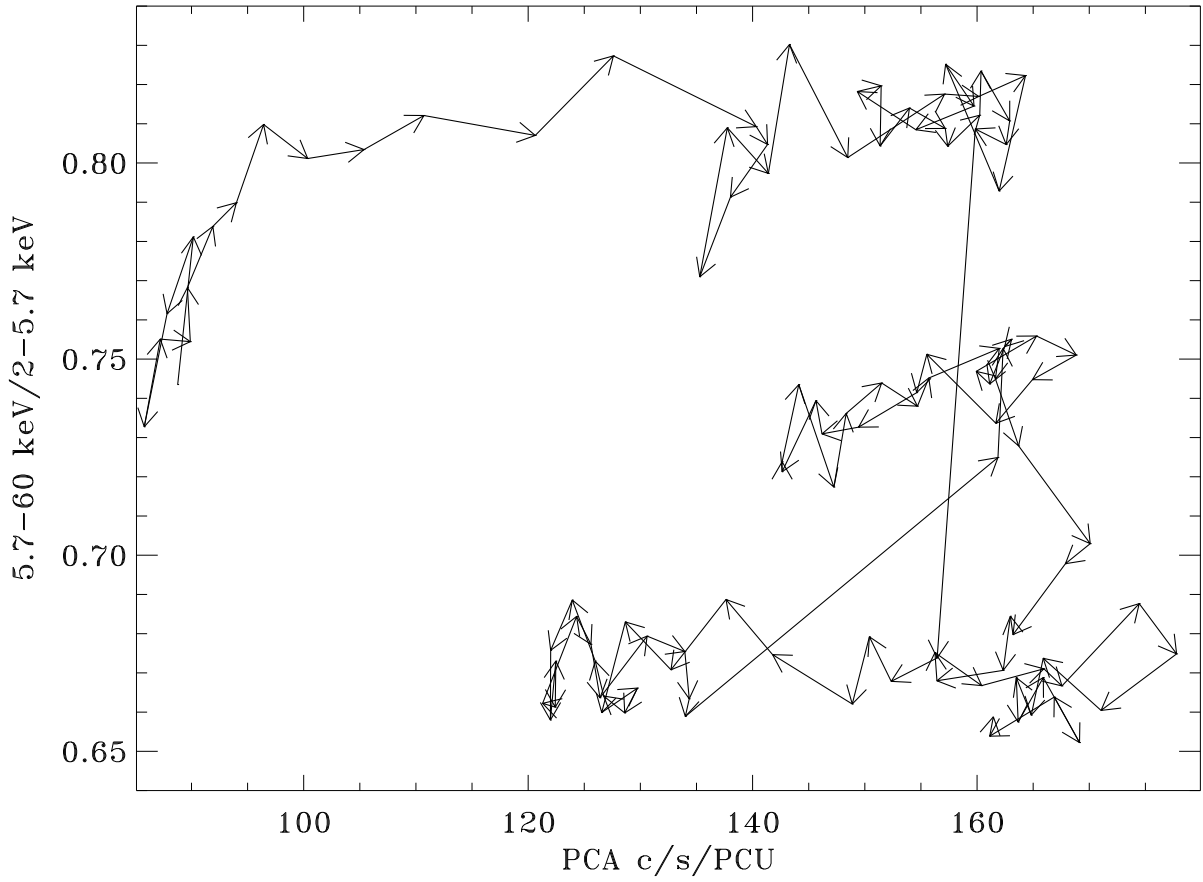,width=5in,angle=90}
\caption{Spectral hysteresis. The figure shows an expanded view of Panel c
in Fig.~7, but now indicates the time sequence of the evolution.  }
\end{figure}
\begin{figure}
\psfig{figure=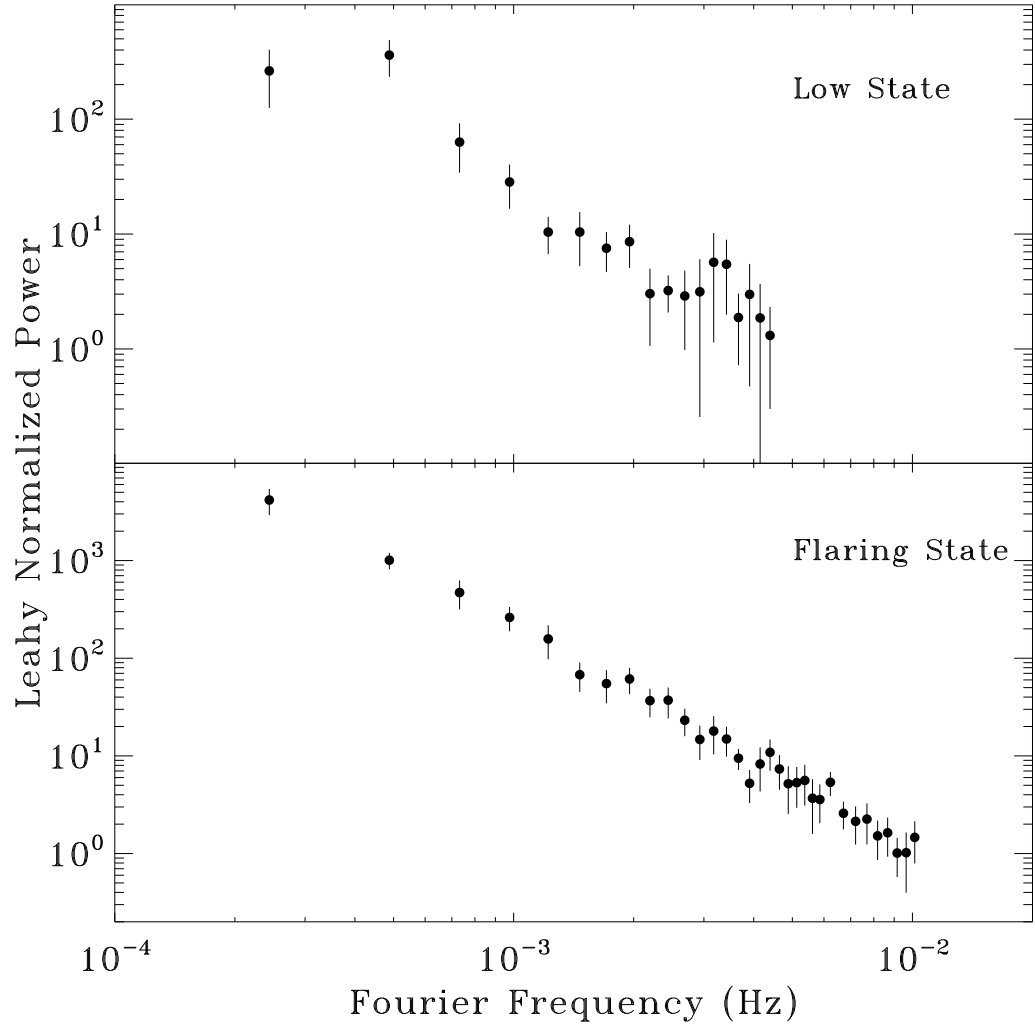,width=6in}
\caption{Power density spectra of Mrk~421. They were constructed from data
with 1/8 s time resolution (see text). }
\end{figure}
\begin{figure}
\psfig{figure=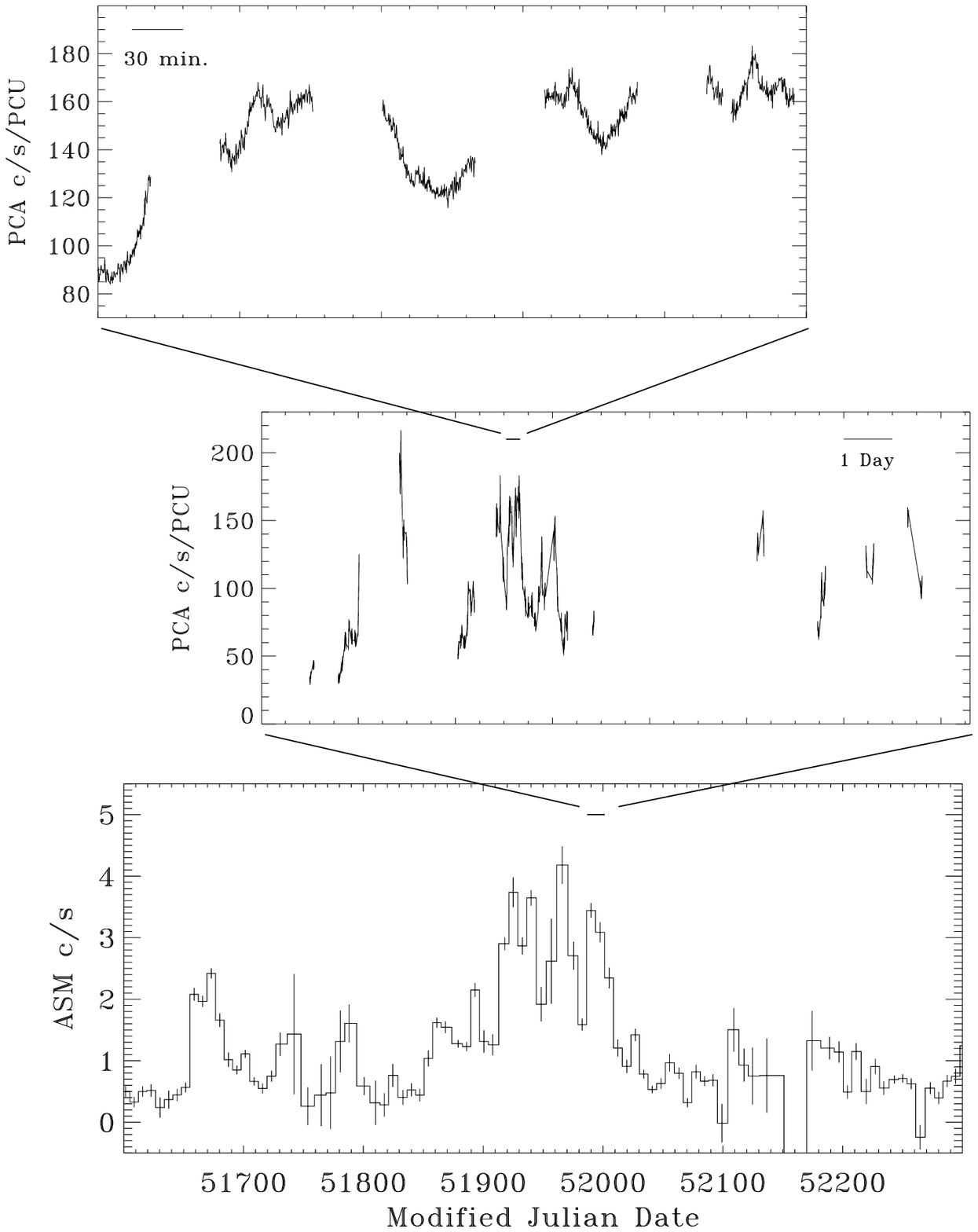,width=6in}
\caption{Hierarchical structure of the X-ray flaring phenomenon in Mrk~421. }
\end{figure}


\clearpage

\begin{references}
\reference{}Barth,~A.~J., Ho,~J.~C., \& Sargent,~W.~L.~W. 2003, ApJ, 583, 134
\reference{}B\"ottcher, M., et al. 1997, A\&A, 324, 395 
\reference{}B\"ottcher,~M., \& Chiang,~J. 2002, ApJ, 581, 127
\reference{}Bradt,~H.~V., Rothschild,~R.~E., \& Swank,~J.~H. 1993, \aaps, 97, 355
\reference{}Brinkmann,~W., Papadakis,~I.~E., A den Herder,~J.~W., \& 
Haberl,~F. 2003, A\&A, 402, 929
\reference{}Buckley,~J.~H., et al. 1996, ApJ, L9
\reference{}Catanese,~M., \& Sambruna,~R.~M. 2000, ApJ, 534, L39
\reference{}Celotti,~A., Fabian,~A.~C., \& Rees,~M.~J. 1998, MNRAS, 293, 239
\reference{}Cui,~W., Feng,~Y.~X., \& Ertmer,~M. 2002, ApJ, 564, L77
\reference{}Dermer,~C.~D. 1998, ApJ, 501, L157
\reference{}Falcone,~A.~D., Cui,~W., \& Finley,~J.~P. 2004, ApJ, in press
\reference{}Fossati,~G., et al. 2000, ApJ, 541, 166
\reference{}Gaidos,~J., et al. 1996, Nature, 383, 319
\reference{}Georganopoulos,~M., \& Kazanas,~D. 2003, ApJ, 594, L27
\reference{}Giebels,~B., et al. 2002, ApJ, 571, 763
\reference{}Kataoka,~J., et al. 2000, ApJ, 528, 243
\reference{}Kataoka,~J., et al. 2001, ApJ, 560, 659
\reference{}Kirk,~J.~G, \& Mastichiadis,~A. 1999, Astropart. Phys., 11, 45
\reference{}Krawczynski,~H., et al. 2001, ApJ, 559, 187
\reference{}Leahy,~D.~A., et al. 1983, \apj, 266, 160
\reference{}Li,~H., \& Kusunose,~M. 2000, ApJ, 536, 729
\reference{}Lu,~E.~T., \& Hamilton,~R.~J. 1991, ApJL, 380, L89
\reference{}Lyutikov,~M. 2003, New Astr. Rev. 47, 513
\reference{}Lyutikov,~M., Pariev, V. I., \& Blandford, R. D. 2003, ApJ, 597, 998
\reference{}Maraschi,~L., et al. 1999, ApJ, 526, L81
\reference{}Mastichiadis, A., \& Kirk, J. G. 1997, A\&A, 320, 19 
\reference{}Mineshige, S., Takeuchi, M., \& Nishimori, H. 1994, ApJ, 435, L125
\reference{}Kato,~S., Fukue,~J., \& Mineshige,~S. 1998, Black-Hole 
Accretion Disks (Kyoto University Press)
\reference{}Protheroe,~R.~J. 2002, PASA, 19, 486
\reference{}Punch,~M., et al. 1992, Nature, 358, 477
\reference{}Rees, M. J. 1978, MNRAS, 184, P61 
\reference{}Rybicki,~G.~B., \& Lightman,~A.~P., Radiative Processes in 
Astrophysics (New York: John Wiley \& Sons)
\reference{}Spada, M., et al. 2001, MNRAS, 325, 1559 
\reference{}Takahashi,~T., et al. 1996, ApJ, 470, L89
\reference{}Tanihata,~C., et al. 2001, ApJ, 563, 569
\reference{}Tanihata, C., Takahashi, T., Kataoka, J., \& Madejski,~G.~M. 2003,
ApJ, 584, 153
\reference{}Urry,~C.~M., \& Padovani,~P. 1995, PASP, 107, 803
\reference{}Zhang,~Y,~H. 2000, MNRAS, 337, 609

\end{references}
\end{document}